\documentstyle[12pt]{article}

\begin{document}

\title{\bf Causal Scattering Matrix and the Chronological Product}

\author{Yury M. Zinoviev\thanks{This work was supported in part by the Russian Foundation
 for Basic Research and the Program for Supporting Leading
 Scientific Schools (Grant No. 8265.2010.1).}}

\date{}
\maketitle

Steklov Mathematical Institute, Gubkin St. 8, 119991, Moscow,
Russia,

e - mail: zinoviev@mi.ras.ru

\vskip 1cm

\noindent {\bf Abstract.} A causal scattering matrix is constructed
by means of the mixed chronological and normal product of the free
quantum fields of different variables $x \in {\bf R}^{4}$. This
scattering matrix does not contain the diverging integrals.

\vskip 1cm

\section{Introduction}
\setcounter{equation}{0}

The scattering matrix connects the asymptotic Schr\"odinger equation
solutions. Following Stueckelberg and Rivier \cite{6} Bogoliubov
\cite{8} introduced the scattering matrix without making use of
Schr\"odinger equation. Bogoliubov \cite{8} defined the function
$g(x)$ taking the values in the interval $[0,1]$ and representing
the intensity of interaction switching. Then in the space-time
domains where $g(x) = 0$ the interaction is absent, in the
space-time domains where $g(x) = 1$ it is switched on absolutely and
for $0 < g(x) < 1$ it is switched on partially. Now let $g(x)$ be
not zero only in some finite space-time domain. In this case the
fields are free in the sufficiently long ago past and in the
sufficiently distant future. Bogoliubov \cite{8} believed that the
initial and final states should be connected by some operator
$S(g)$. The operator $S(g)$ is naturally interpreted as the
scattering operator for the case when the interaction is switched on
with the intensity $g(x)$. Bogoliubov \cite{8} believed also that
the "physical" case when the interaction is switched on absolutely
in the whole space-time must be considered in the given scheme by
making use of the limit process when the space-time domain where
$g(x) = 1$ spreads infinitely to the whole space-time. If for some
matrix elements of the operator $S(g)$ the limit values exist, then
these limit matrix elements ought be considered as the corresponding
matrix elements of the scattering matrix $S$. The mathematical
reason for the switching function $g(x)$ is very simple: the
distributions should be integrated with the smooth functions rapidly
decreasing at the infinity.

Let us formulate the main physical conditions the operator $S(g)$
should satisfy. In order to guarantee the theory covariance we need
to demand
\begin{equation}
\label{3.4} S(Lg) = U(L)S(g)U^{\ast}(L)
\end{equation}
where $Lg(x) = g(L^{- 1}x)$ and $U(L)$ is a unitary operator by
means of which the quantum wave functions transform under the
transformations $L$ from the Lorentz group and the group of
translations. In order to conserve the amplitude state norm under
the transformation from the initial state to the final state the
operator $S(g)S^{\ast}(g)$ has to be the projector on the subspace
of the asymptotic states. The operator $S(g)$ is defined on the
subspace of the asymptotic states. Bogoliubov \cite{8} required that
the operator $S(g)$ satisfies the unitary condition:
\begin{equation}
\label{3.5} S(g)S^{\ast}(g) = 1.
\end{equation}
The identity operator is denoted by $1$ and is often omitted.

Now let us take into account the causality condition according to
which some event in the system can influence the evolution of the
system in the future only and can not influence the behavior of the
system in the past, in the time preceding the given event. Therefore
we need to demand that a change in the interaction law in some
space-time domain can change the motion in the succeeding moments
only. Due to the book (\cite{1}, Section 17.5) we formulate the
causality condition. We consider the case when the space-time domain
$G$ where the function $g(x)$ is not zero is divided into two
separate domains $G_{1}$ and $G_{2}$ such that all time points of
the domain $G_{1}$ lie in the past relative to some moment $t$ and
all time points of the domain $G_{2}$ lie in the future relative to
$t$. Then the function $g(x)$ may be represented as a sum of two
functions
\begin{equation}
\label{3.6} g(x) = g_{1}(x) + g_{2}(x)
\end{equation}
where the function $g_{1}$ is not zero in the domain $G_{1}$ only
and the function $g_{2}$ is not zero in the domain $G_{2}$ only. The
causality condition is called the relation
\begin{equation}
\label{3.9} S(g_{1} + g_{2}) = S(g_{2})S(g_{1}).
\end{equation}
Bogoliubov \cite{8} defines the causal scattering matrix operator
$S(g)$ by means of Lagrange function constructed from the normal
products of the free quantum fields of one variable $x \in {\bf
R}^{4}$. The causal scattering matrix operator $S(g)$ coefficients
contain the diverging integrals.

In order to "quantize" the physicists change the powers of the
classical variables in the classical Lagrange function for the
normal products of the free field operators at the same space-time
point. The "quantized" Lagrangian mechanics is not compatible with
the causality condition (\ref{3.9}).

This paper is the straightforward generalization of the paper
\cite{8}. In this paper Lagrange function is changed for a linear
combination of the normal products of the free quantum fields of
different variables $x \in {\bf R}^{4}$. The switching function
$g(x)$ of one variable $x \in {\bf R}^{4}$ is changed for the set of
the switching functions of different variables $x \in {\bf R}^{4}$.
The causal scattering matrix operator satisfying the relation of the
type (\ref{3.9}) is constructed by means of the mixed chronological
and normal product of the free quantum fields of different variables
$x \in {\bf R}^{4}$. The causal scattering matrix coefficients do
not contain the diverging integrals.

\section{Chronological Product}
\setcounter{equation}{0}

Consider a free real scalar field and a free spin field given by the
distributions $\varphi (x)$ and $\psi_{\alpha}(x)$ taking the values
in the set of Hilbert space operators with the commutation relations
(11.3) and (13.4) from the book \cite{1}
\begin{equation}
\label{2.1} [\varphi (x),\varphi (y)] = \frac{1}{i} D_{m^{2}}(x -
y),
\end{equation}
$$
[\psi_{\alpha} (x),\bar{\psi_{\beta}} (y)]_{+} \equiv \psi_{\alpha}
(x)\bar{\psi_{\beta}} (y) + \bar{\psi_{\beta}} (y)\psi_{\alpha} (x)
= \frac{1}{i} \left( i\sum_{\mu \, =\, 0}^{3} \gamma_{\alpha
\beta}^{\mu} \frac{\partial}{\partial x^{\mu}} + m\right)
D_{m^{2}}(x - y)
$$
where the Pauli - Jordan distribution (\cite{1}, relation (10.18))
\begin{equation}
\label{2.2} D_{m^{2}}(x) = \frac{i}{(2\pi)^{3}} \int d^{4}k (\theta
(k^{0})\delta ((k,k) - m^{2}) - \theta (- k^{0})\delta ((k,k) -
m^{2})) e^{- i(k,x)},
\end{equation}
$$
(k,x) = \sum_{\mu, \nu \, =\, 0}^{3} \eta_{\mu \nu} k^{\mu}x^{\nu}.
$$
Here $\eta_{\mu \nu}$ is the diagonal $4\times 4$ - matrix with the
diagonal elements $ \eta_{00} = - \eta_{11} = - \eta_{22} = -
\eta_{33} = 1$ and $\gamma_{\alpha \beta}^{\mu}$ is the Dirac
matrices (\cite{1}, relations (6.18)). The commuting relations of
the free vector field $U_{\mu}(x)$ and the free electromagnetic
field $A_{\mu}(x)$ are given by the relations (11.27) and (12.4)
from the book \cite{1}. These relations are similar to the relations
(\ref{2.1}). The operator valued distributions $\varphi (x)$,
$U_{\mu}(x)$, $A_{\mu}(x)$, $\psi_{\alpha}(x)$ and all its possible
derivatives are called the free quantum fields and denote
$u_{\alpha}(x)$.

The vacuum expectation of the product of two free fields is given by
the relations (10.17), (16.12) and (16.14) from the book \cite{1}
\begin{equation}
\label{2.3} <\varphi (x)\varphi (y)>_{0} = \frac{1}{i}
D_{m^{2}}^{-}(x - y) \equiv \frac{1}{(2\pi)^{3}} \int d^{4}k \theta
(k^{0})\delta ((k,k) - m^{2})e^{- i(k,x - y)},
\end{equation}
$$
<\psi_{\alpha} (x)\bar{\psi}_{\beta} (y)>_{0} = \frac{1}{i} \left(
i\sum_{\mu \, =\, 0}^{3} \gamma_{\alpha \beta}^{\mu}
\frac{\partial}{\partial x^{\mu}} + m\right) D_{m^{2}}^{-}(x - y).
$$
The vacuum expectations $<U_{\lambda}^{\ast}(x)U_{\nu}(y)>_{0}$ and
$<A_{\lambda}(x)A_{\nu}(y)>_{0}$ are similar to the vacuum
expectations (\ref{2.3}). The vacuum expectations of another free
fields products are either the derivatives of the distributions
(\ref{2.3}) or are equal to zero.

The free fields normal product is given by the relations (16.17)
from the book \cite{1}
$$
:1:\, =\, 1,\, \, :u_{\alpha} (x):\, =\, u_{\alpha} (x),
$$
\begin{eqnarray}
\label{2.4} u_{\alpha (1)} (x_{1}) \cdots u_{\alpha (n)} (x_{n}) =
:u_{\alpha (1)} (x_{1}) \cdots u_{\alpha (n)} (x_{n}): + \sum_{1\,
\leq \, k\, <\, l\, \leq
\, n} <u_{\alpha (k)} (x_{k}) u_{\alpha (l)} (x_{l})>_{0} \times \nonumber \\
:u_{\alpha (1)} (x_{1}) \cdots \widehat{u_{\alpha (k)} (x_{k})}
\cdots \widehat{u_{\alpha (l)} (x_{l})} \cdots u_{\alpha (n)}
(x_{n}): + \cdots, \, \, n = 2,3,...
\end{eqnarray}
The subsequent summings in the equality (\ref{2.4}) run over two
pairs of numbers from $1,...,n$, over three pairs of numbers from
$1,...,n$, etc. In the book ((\cite{1}), Section 16.2) the
definition (\ref{2.4}) is called the Wick theorem for the normal
products. The normal product may be also defined in the following
way
\begin{eqnarray}
\label{2.5} :u_{\alpha (1)} (x_{1}) \cdots u_{\alpha (n)} (x_{n}): =
u_{\alpha (1)} (x_{1}) \cdots u_{\alpha (n)} (x_{n}) - \sum_{1\,
\leq \, k\, <\, l\, \leq
\, n} <u_{\alpha (k)} (x_{k}) u_{\alpha (l)} (x_{l})>_{0} \times \nonumber \\
u_{\alpha (1)} (x_{1}) \cdots \widehat{u_{\alpha (k)} (x_{k})}
\cdots \widehat{u_{\alpha (l)} (x_{l})} \cdots u_{\alpha (n)}
(x_{n}) + \cdots, \, \, n = 2,3,...
\end{eqnarray}
In the equality (\ref{2.5}) the subsequent summing run over two
pairs of numbers from $1,..,n$, over three pairs of numbers from
$1,..,n$, etc. The summing over an even (odd) number of pairs has
the sign plus (minus). The relation (\ref{2.5}) for $n = 2$
coincides with the relation (\ref{2.4}) for $n = 2$.

Let us prove the relation (\ref{2.5}) by making use of the relation
(\ref{2.4}). Let us change every distribution $<u_{\alpha (k)}
(x_{k}) u_{\alpha (l)} (x_{l})>_{0}$ in the right-hand side of the
relation (\ref{2.4}) for the distribution
\begin{equation}
\label{2.51} <u_{\alpha (k)} (x_{k}) u_{\alpha (l)} (x_{l})>_{0}  -
<u_{\alpha (k)} (x_{k}) u_{\alpha (l)} (x_{l})>_{0}
\end{equation}
equal to zero. Then we get the relation
\begin{eqnarray}
\label{2.52} :u_{\alpha (1)} (x_{1}) \cdots u_{\alpha (n)} (x_{n}):
= :u_{\alpha (1)} (x_{1}) \cdots u_{\alpha (n)} (x_{n}): + \nonumber
\\ \sum_{1\, \leq \, k\, <\, l\, \leq \, n} (<u_{\alpha (k)} (x_{k})
u_{\alpha (l)} (x_{l})>_{0}  -
<u_{\alpha (k)} (x_{k}) u_{\alpha (l)} (x_{l})>_{0}) \times \nonumber \\
:u_{\alpha (1)} (x_{1}) \cdots \widehat{u_{\alpha (k)} (x_{k})}
\cdots \widehat{u_{\alpha (l)} (x_{l})} \cdots u_{\alpha (n)}
(x_{n}): + \cdots
\end{eqnarray}
Choose the first term $<u_{\alpha (k)} (x_{k}) u_{\alpha (l)}
(x_{l})>_{0}$ in every sum (\ref{2.51}) of the equality
(\ref{2.52}). Adding the first term $:u_{\alpha (1)} (x_{1}) \cdots
u_{\alpha (n)} (x_{n}):$ we get due to the relation (\ref{2.4}) the
first term of the right-hand side of the relation (\ref{2.5}). Let
us choose the second term $ - <u_{\alpha (k)} (x_{k}) u_{\alpha (l)}
(x_{l})>_{0}$ in one sum (\ref{2.51}) of the relation (\ref{2.52})
and the first term $<u_{\alpha (k)} (x_{k}) u_{\alpha (l)}
(x_{l})>_{0}$ in all other sums (\ref{2.51}). Due to the relation
(\ref{2.4}) we get the second term of the right-hand side of the
relation (\ref{2.5}). If we continue this process, we transform the
relation (\ref{2.52}) into the relation (\ref{2.5}).

Let the vacuum expectation of normal product of arbitrary number $n
> 0$ of free fields vanish. Let also $<1>_{0} = 1$. Hence the
relation (\ref{2.4}) implies that the vacuum expectation of any odd
number of free fields vanishes and the vacuum expectation of any
even number of free fields is equal to
\begin{eqnarray}
\label{2.6} <u_{\alpha (1)} (x_{1}) \cdots u_{\alpha (2n)}
(x_{2n})>_{0} = \nonumber \\ \sum_{\sigma} <u_{\alpha (\sigma (1))}
(x_{\sigma (1)}) u_{\alpha (\sigma (2))} (x_{\sigma (2)})>_{0}
\cdots <u_{\alpha (\sigma (2n - 1))} (x_{\sigma (2n - 1)}) u_{\alpha
(\sigma (2n))} (x_{\sigma (2n)})>_{0}
\end{eqnarray}
where the summing runs over all permutations $\sigma$ of the numbers
$1,...,2n$  not changing the order in any pair of the numbers $2k -
1$, $2k$: $\sigma (2k - 1) < \sigma (2k)$ for any $k = 1,...,n$.

By making use of the relations (\ref{2.5}), (\ref{2.6}) we get the
rule for calculation of the vacuum expectation
$$
<u_{\beta (1)} (y_{1}) \cdots u_{\beta (m)} (y_{m}) :u_{\alpha (1)}
(x_{1}) \cdots u_{\alpha (n)} (x_{n}): u_{\beta^{\prime} (1)}
(y_{1}^{\prime}) \cdots u_{\beta^{\prime} (m^{\prime})}
(y_{m^{\prime}}^{\prime})>_{0}.
$$
In the sum (\ref{2.6}) for the vacuum expectation
\begin{equation}
\label{2.600} <u_{\beta (1)} (y_{1}) \cdots u_{\beta (m)}(y_{m})
u_{\alpha (1)} (x_{1}) \cdots u_{\alpha (n)} (x_{n})
u_{\beta^{\prime} (1)} (y_{1}^{\prime}) \cdots u_{\beta^{\prime}
(m^{\prime})} (y_{m^{\prime}}^{\prime})>_{0}
\end{equation}
it is needed to cancel all terms containing at least one multiplier
$<u_{\alpha (k)} (x_{k}) u_{\alpha (l)} (x_{l})>_{0}$, $1 \leq k < l
\leq n$. Hence it is possible to let $x_{1} = \cdots = x_{n}$ in the
vacuum expectation
$$
<u_{\beta (1)} (y_{1}) \cdots u_{\beta (m)} (y_{m}) :u_{\alpha (1)}
(x_{1}) \cdots u_{\alpha (n)} (x_{n}): u_{\beta^{\prime} (1)}
(y_{1}^{\prime}) \cdots u_{\beta^{\prime} (m^{\prime})}
(y_{m^{\prime}}^{\prime})>_{0}.
$$
Therefore there exists the integral
\begin{equation}
\label{2.60} \int d^{4}x_{1} \cdots d^{4}x_{n} :u_{\alpha (1)}
(x_{1}) \cdots u_{\alpha (n)} (x_{n}): h(x_{1},..,x_{n})
\end{equation}
for the distribution
\begin{equation}
\label{2.601} h(x_{1},..,x_{n}) = g(x_{1})\delta (x_{2} - x_{1})
\cdots \delta (x_{n} - x_{1}),\, \, g(x_{1}) \in D({\bf R}^{4}).
\end{equation}
The integral (\ref{2.60}),  (\ref{2.601}) exists for free fields
only. The normal product of interacting fields is not defined. It is
impossible to let $x_{1} = \cdots = x_{n}$ in the expectation
(\ref{2.600}) of the free fields product $u_{\alpha (1)} (x_{1})
\cdots u_{\alpha (n)} (x_{n})$. The integral (\ref{2.60}),
(\ref{2.601}) we denote as
\begin{equation}
\label{2.61} \int d^{4}x :u_{\alpha (1)} (x) \cdots u_{\alpha (n)}
(x): g(x).
\end{equation}
If Fermi fields (the operators $\psi_{\alpha} (x)$,
$\bar{\psi_{\alpha}} (x)$ and their derivatives) are included in the
normal product (\ref{2.60}) in the even combinations, the operator
(\ref{2.60}) is called polylocal (\cite{1}, Section 16.8). The
polylocal operator (\ref{2.61}) is called local (\cite{1}, Section
16.8). Due to the book (\cite{1}, Section 18.3) "the interaction
Lagrangian should be the local, Hermitian and Lorentz covariant
combination of field operator functions."  We consider the polylocal
combination of the field operator functions (\ref{2.60}). If we take
the distribution (\ref{2.601}), we get the local interaction
Lagrangian (\ref{2.61}).

The chronological product of the field operators is defined by the
relation (19.1) from the book \cite{1}
\begin{equation}
\label{2.7} T(u_{\alpha (1)}(x_{1}); \cdots ;u_{\alpha (n)}(x_{n}))
= (- 1)^{p}u_{\alpha (j_{1})}(x_{j_{1}}) \cdots u_{\alpha
(j_{n})}(x_{j_{n}}),\, \, x_{j_{1}}^{0} \geq x_{j_{2}}^{0} \geq
\cdots \geq x_{j_{n}}^{0}
\end{equation}
where $p$ is the parity of the Fermi fields permutation
corresponding to the permutation $j$ transforming the numbers
$1,2,...,n$ into the numbers $j_{1},j_{2},...,j_{n}$. Due to the
paper \cite{8}: "Let us note as Stueckelberg did that the usual
definition of $T$ - product by means of introduction the
chronological order for the operators is effective only without the
coincidence of the arguments $x_{1},...,x_{n}$. In view of the
corresponding coefficient functions singularity their "redefinition"
in the domains of the arguments coincidence is not done explicitly
and presents a special problem."

\noindent It is necessary to change the closed set $x_{j_{1}}^{0}
\geq x_{j_{2}}^{0} \geq \cdots \geq x_{j_{n}}^{0}$ for the open set
$x_{j_{1}}^{0} > x_{j_{2}}^{0} > \cdots > x_{j_{n}}^{0}$ in the
definition (\ref{2.7}). The distribution may be restricted only to
the open set. The correct relation (\ref{2.7}) does not define the
chronological product in the domains of the time arguments
coincidence.

In the book (\cite{1}, Section 19.2) another definition of the field
operators chronological product for $n = 2,3,...$ is obtained
\begin{eqnarray}
\label{2.8} T(u_{\alpha (1)} (x_{1}); \cdots ;u_{\alpha (n)}
(x_{n})) = :u_{\alpha (1)} (x_{1}) \cdots u_{\alpha (n)} (x_{n}): +
\sum_{1\, \leq \, k\, <\, l\, \leq \, n} \nonumber \\ <T(u_{\alpha
(k)} (x_{k}); u_{\alpha (l)} (x_{l}))>_{0} :u_{\alpha (1)} (x_{1})
\cdots \widehat{u_{\alpha (k)} (x_{k})} \cdots \widehat{u_{\alpha
(l)} (x_{l})} \cdots u_{\alpha (n)} (x_{n}): + \cdots
\end{eqnarray}
In the following terms in the equality (\ref{2.8}) the summings run
over two pairs of the numbers from $1,..,n$, over three pairs, etc.
Due to the relations (19.6) and (19.9) from the book \cite{1}
\begin{equation}
\label{2.9} <T(\varphi (x);\varphi (y))>_{0} = \frac{1}{i}
D_{m^{2}}^{c}(x - y) \equiv \lim_{\epsilon \, \rightarrow \, + 0}
\frac{1}{(2\pi)^{4}i} \int d^{4}k \frac{e^{i(k,x - y)}}{m^{2} -
(k,k) - i\epsilon},
\end{equation}
$$
<T(\psi_{\alpha} (x);\bar{\psi}_{\beta} (y))>_{0} = \frac{1}{i}
\left( i\sum_{\mu \, =\, 0}^{3} \gamma_{\alpha \beta}^{\mu}
\frac{\partial}{\partial x^{\mu}} + m\right) D_{m^{2}}^{c}(x - y).
$$
The vacuum expectations $<T(A_{\lambda}(x);A_{\nu}(y))>_{0}$ and
$<T(U_{\lambda}^{\ast}(x);U_{\nu}(y))>_{0}$ are similar to the
vacuum expectations (\ref{2.9}). The distributions $<T(u_{\alpha
(1)}(x_{1})u_{\alpha (2)}(x_{2}))>_{0}$ for other free fields are
the derivatives of the distributions (\ref{2.9}) or are equal to
zero. In the book (\cite{1}, Section 19.2) the relation (\ref{2.8})
is called the Wick theorem for chronological products. If we replace
the distributions $<u_{\alpha (k)} (x_{k}) u_{\alpha (l)}
(x_{l})>_{0}$ with the distributions $<T(u_{\alpha (k)} (x_{k});
u_{\alpha (l)} (x_{l}))>_{0}$ in the relation (\ref{2.4}), then we
get the relation (\ref{2.8}).

Due to the relation (14.12) from the book \cite{1}
\begin{equation}
\label{2.91} D_{m^{2}}^{c}(x) = \left\{ {D_{m^{2}}^{-}(x), \hskip
0,5cm x^{0}
> 0,} \atop  {D_{m^{2}}^{-}(- x), \hskip 0,3cm x^{0} < 0.}\right.
\end{equation}
Hence the distribution $D_{m^{2}}^{c}(x) - D_{m^{2}}^{-}(x) = 0$ for
$x^{0} > 0$. The definitions (\ref{2.3}), (\ref{2.9}) imply Lorentz
invariance of the distribution $D_{m^{2}}^{c}(x) -
D_{m^{2}}^{-}(x)$. Hence the support of this distribution lies in
the closed lower light cone.  The distribution $D_{m^{2}}^{-}(x)$
satisfies the Klein - Gordon equation
\begin{equation}
\label{2.10} ((\partial_{x}, \partial_{x}) + m^{2})D_{m^{2}}^{-}(x)
= 0, \, \, (\partial_{x}, \partial_{x}) = \sum_{\mu \, =\, 0}^{3}
\eta_{\mu \mu} \left( \frac{\partial}{\partial x^{\mu}}\right)^{2}.
\end{equation}
The distribution $D_{m^{2}}^{c}(x)$ is the fundamental solution of
the Klein - Gordon equation
\begin{equation}
\label{2.11} ((\partial_{x}, \partial_{x}) + m^{2}) D_{m^{2}}^{c}(x)
= \delta (x).
\end{equation}
Hence the distribution $D_{m^{2}}^{c}(x) - D_{m^{2}}^{-}(x)$
satisfies the equation (\ref{2.11}). Let us prove that the equation
(\ref{2.11}) has the unique solution in the class of the
distributions with supports in the closed lower light cone. Let the
equation (\ref{2.11}) have two solutions $e^{(1)}(x)$, $e^{(2)}(x)$
with supports  in the closed lower light cone. Since its supports
lie in the closed lower light cone, the convolution is defined. The
convolution commutativity implies the coincidence of these solutions
\begin{eqnarray}
\label{2.12} e^{(2)}(x) = ((\partial_{x}, \partial_{x}) + m^{2})
\int d^{4}ye^{(1)}(x - y)e^{(2)}(y) = \nonumber \\ ((\partial_{x},
\partial_{x}) + m^{2}) \int d^{4}ye^{(2)}(x - y)e^{(1)}(y) =
e^{(1)}(x).
\end{eqnarray}
Therefore the distribution $e^{(1)}(x)$ coincides with the
distribution
\begin{equation}
\label{2.13} D_{m^{2}}^{ret}(- x) = \lim_{\epsilon \, \rightarrow \,
+\, 0} \frac{1}{(2\pi)^{4}} \int d^{4}k \frac{e^{i(k,x)}}{m^{2} -
(k^{0} + i\epsilon)^{2} + |{\bf k}|^{2}}
\end{equation}
given by the relation (14.7) from the book \cite{1} and
\begin{equation}
\label{2.15} D_{m^{2}}^{c}(x) = D_{m^{2}}^{ret}(- x) +
D_{m^{2}}^{-}(x).
\end{equation}
The distribution (\ref{2.13}) is fundamental for the quantum field
theory. It seems natural to use the special notation instead the
cumbersome notation $- D_{m^{2}}^{ret}(- x)$ of the book \cite{1}.

The substitution of the equality (\ref{2.15}) into the relations
(\ref{2.9}) yields
\begin{equation}
\label{2.16} <T(u_{\alpha (1)}(x_{1});u_{\alpha (2)}(x_{2}))>_{0} =
<u_{\alpha (1)}(x_{1})u_{\alpha (2)}(x_{2})>_{0} + <u_{\alpha
(1)}(x_{1})u_{\alpha (2)}(x_{2})>_{c},
\end{equation}
\begin{equation}
\label{2.17} <\varphi(x)\varphi(y)>_{c} = \frac{1}{i}
D_{m^{2}}^{ret}(y - x),
\end{equation}
$$
<\psi_{\alpha} (x)\bar{\psi_{\beta}} (y)>_{c} = \frac{1}{i} \left(
i\sum_{\mu \, =\, 0}^{3} \gamma_{\alpha \beta}^{\mu}
\frac{\partial}{\partial y^{\mu}} + m\right) D_{m^{2}}^{ret}(y - x).
$$
The distributions $<U_{\lambda}^{\ast}(x)U_{\nu}(y)>_{c}$ and
$<A_{\mu}(x)A_{\nu}(y)>_{c}$ are similar to the distributions
(\ref{2.17}). The distributions $<u_{\alpha (1)}(x_{1})u_{\alpha
(2)}(x_{2})>_{c}$ for other free fields are the derivatives of the
distributions (\ref{2.17}) or are equal to zero.

We substitute the relation (\ref{2.16}) into the right-hand side of
the equality (\ref{2.8}). Let us take the distribution $<u_{\alpha
(1)}(x_{1})u_{\alpha (2)}(x_{2})>_{0}$ in any sum (\ref{2.16}) in
the equality (\ref{2.8}). Then we take the distribution $<u_{\alpha
(1)}(x_{1})u_{\alpha (2)}(x_{2})>_{c}$ in one sum (\ref{2.16}) in
the equality (\ref{2.8}), take the distribution $<u_{\alpha
(1)}(x_{1})u_{\alpha (2)}(x_{2})>_{0}$ in all other sums
(\ref{2.16}) in the right-hand side of the equality (\ref{2.8}) and
so on. Therefore we have
\begin{eqnarray}
\label{2.19} T(u_{\alpha (1)}(x_{1});\cdots ;u_{\alpha (n)}(x_{n}))
=  \nonumber
\\ \Biggl\{ :u_{\alpha (1)} (x_{1}) \cdots u_{\alpha (n)} (x_{n}): +
\sum_{1\, \leq \, k\, <\, l\, \leq
\, n} <u_{\alpha (k)} (x_{k}) u_{\alpha (l)} (x_{l})>_{0} \times \nonumber \\
:u_{\alpha (1)} (x_{1}) \cdots \widehat{u_{\alpha (k)} (x_{k})}
\cdots \widehat{u_{\alpha (l)} (x_{l})} \cdots u_{\alpha (n)}
(x_{n}): + \cdots \Biggr\} + \nonumber \\ \sum_{1\, \leq \, k\, <\,
l\, \leq \, n} <u_{\alpha (k)}(x_{k})u_{\alpha (l)}(x_{l})>_{c}
\times \nonumber
\\ \Biggl\{ :u_{\alpha
(1)}(x_{1}) \cdots \widehat{u_{\alpha (k)}(x_{k})} \cdots
\widehat{u_{\alpha (l)}(x_{l})} \cdots u_{\alpha (n)}(x_{n}): +
\cdots \Biggr\} + \cdots.
\end{eqnarray}
The following summings run over two pairs of the numbers from
$1,..,n$, over three pairs of the numbers, etc. By making use of the
relation (\ref{2.4}) it is possible to rewrite the equality
(\ref{2.19}) as
\begin{eqnarray}
\label{2.18} T(u_{\alpha (1)}(x_{1});\cdots ;u_{\alpha (n)}(x_{n}))
= u_{\alpha (1)}(x_{1}) \cdots u_{\alpha (n)}(x_{n}) + \nonumber \\
\sum_{1\, \leq \, k\, <\, l\, \leq \, n} <u_{\alpha
(k)}(x_{k})u_{\alpha (l)}(x_{l})>_{c} \times \nonumber \\ u_{\alpha
(1)}(x_{1}) \cdots \widehat{u_{\alpha (k)}(x_{k})} \cdots
\widehat{u_{\alpha (l)}(x_{l})} \cdots u_{\alpha (n)}(x_{n})  +
\cdots.
\end{eqnarray}
The following summings run over two pairs of the numbers from
$1,..,n$, over three pairs of the numbers, etc. The relations
(\ref{2.17}), (\ref{2.18}) yield the definition of the chronological
product of free field operators. It is sufficient to use the
distributions (\ref{2.17}) for the definition (\ref{2.18}) of the
chronological product. The distributions (\ref{2.3}) and (\ref{2.9})
are needed for the definition (\ref{2.8}) of the chronological
product.

The difference between the chronological product of the free quantum
field operators $T(u_{\alpha (1)}(x_{1});\cdots ;u_{\alpha
(n)}(x_{n}))$ and the usual product of these fields $u_{\alpha
(1)}(x_{1})\cdots u_{\alpha (n)}(x_{n})$ is represented by the sum
of the terms proportional to the distributions $<u_{\alpha
(k)}(x_{k})u_{\alpha (l)} (x_{l})>_{c}$, $1 \leq k < l \leq n$. The
distribution $D_{m^{2}}^{ret}(y - x)$ vanishes except for the
vectors $x - y$ lying in the closed lower light cone. Thus the
chronological product of the free quantum field operators
$T(u_{\alpha (1)}(x_{1});\cdots ;u_{\alpha (n)}(x_{n}))$ differs
from the usual product of these free quantum field operators
$u_{\alpha (1)}(x_{1})\cdots u_{\alpha (n)}(x_{n})$ in the only case
when for some numbers $1 \leq k < l \leq n$ the argument difference
$x_{k} - x_{l}$ lies in the closed lower light cone. In the
definition of the chronological product the distributions
$<u_{\alpha (k)}(x_{k})u_{\alpha (l)} (x_{l})>_{c}$, $1 \leq k < l
\leq n$ define the delays:
$$
D_{0}^{ret}(x) = (2\pi)^{- 1} \theta (x^{0})\delta ((x,x)).
$$
The Newton gravity law requires the instant propagation of the force
action. The special relativity requires that the propagation speed
does not exceed that of light. (We believe that the gravity
propagation speed coincides with the light speed.) It requires also
the gravity laws covariance under Lorentz transformation. Long ago
Poincar\'e \cite{4} tried to find such a modification of the Newton
gravity law: "First of all, it enables us to suppose that the
gravity forces propagate not instantly, but at the light velocity".
The interaction force of two physical points should depend not on
their simultaneous positions and speeds but on the positions and the
speeds at the time moments which differ from each other in the
interval needed for light covering the distance between the physical
points. (The interaction force of two physical points should depend
also on the acceleration of one physical point at the delayed time
moment.) The delay is one of possible causality condition
statements. The Lorentz covariance and the causality condition are
the crucial points of the relativistic quantum field theory. These
conditions were proposed by Poincar\'e \cite{4} for the relativistic
causal gravity law. These conditions should be valid for any
interaction.

The distribution $D_{m^{2}}^{c}(x - y)$ defines the vacuum
expectation of the chronological product (\ref{2.17}), (\ref{2.18})
of two free quantum fields. Due to Stueckelberg and Rivier \cite{6}
the classical "causal action" is given by the distribution
$D_{m^{2}}^{ret}(y - x)$ and the distribution $D_{m^{2}}^{c}(x - y)
= D_{m^{2}}^{c}(y - x)$ defines the probability amplitude of the
"causal action".

Let us prove that the chronological product definition (\ref{2.18})
is in accordance with the correct definition (\ref{2.7}). We rewrite
the definition (\ref{2.18}) in the recurrent way:
$$
T(u_{\alpha} (x)) = u_{\alpha} (x),
$$
\begin{eqnarray}
\label{2.20} T(u_{\alpha (1)}(x_{1}); \cdots ; u_{\alpha (n +
1)}(x_{n + 1})) = T(u_{\alpha (1)}(x_{1}); \cdots ; u_{\alpha
(n)}(x_{n}))u_{\alpha (n + 1)}(x_{n + 1}) + \nonumber
\\ \sum_{k\, =\, 1}^{n} <u_{\alpha (k)}(x_{k})u_{\alpha (n + 1)}(x_{n + 1})>_{c}
T(u_{\alpha (1)}(x_{1}); \cdots ; \widehat{u_{\alpha (k)}(x_{k})};
\cdots ;u_{\alpha (n)}(x_{n})).
\end{eqnarray}
Let us consider the permutation $j_{1},j_{2}$ of the numbers $1,2$.
The relation (\ref{2.1}), the relation (14.8) from the book \cite{1}
\begin{equation}
\label{2.21} D_{m^{2}}^{ret}(x) = \left\{ {D_{m^{2}}(x), \hskip
0,5cm x^{0}
> 0,} \atop  {0, \hskip 1,7cm x^{0} < 0,}\right.
\end{equation}
and the relations (\ref{2.18}) for $n = 2$ imply that the relation
(\ref{2.7}) for $n = 2$ is valid for the chronological product
(\ref{2.18}). Let us consider the permutation $j_{1},...,j_{m}$ of
the numbers $1,...,m$. Suppose that for the coordinates
$x_{j_{1}}^{0} > \cdots > x_{j_{m}}^{0}$ and for any number $m =
2,..,n$ the relation (\ref{2.7}) is valid for the chronological
product (\ref{2.18}). Hence the definition (\ref{2.20}) and the
relations (\ref{2.1}), (\ref{2.21}) imply the relation (\ref{2.7})
for the chronological product (\ref{2.18}) in the case of $n + 1$
fields for any permutation $j_{1},...,j_{n + 1}$ of the numbers
$1,...,n + 1$ and for the coordinates $x_{j_{1}}^{0} > x_{j_{2}}^{0}
> \cdots > x_{j_{n + 1}}^{0}$.

Define the mixed chronological and normal product of the free field
operators. The sum for the chronological product
$$
T\left( :\prod_{i\, =\, 1}^{n_{1}} u_{\alpha (i)}(x_{i}):;\cdots
;:\prod_{i\, =\, n_{1} + \cdots + n_{k - 1} + 1}^{n_{1} + \cdots +
n_{k}} u_{\alpha (i)}(x_{i}):\right)
$$
is the sum (\ref{2.18}) for the chronological product $T(u_{\alpha
(1)}(x_{1}); \cdots ;u_{\alpha (n_{1} + \cdots + n_{k})}(x_{n_{1} +
\cdots + n_{k}}))$ where all distributions $<u_{\alpha (m)}(x_{m})
u_{\alpha (l)}(x_{l})>_{c}$ with the arguments from the same group:
$n_{1} + \cdots + n_{j - 1} < m < l \leq n_{1} + \cdots + n_{j}$ are
replaced by the distributions $ - <u_{\alpha (m)}(x_{m})u_{\alpha
(l)}(x_{l})>_{0}$. Thus the chronological order is introduced only
for the free field operators $u_{\alpha (m)}(x_{m})$, $u_{\alpha
(l)}(x_{l})$ the arguments $x_{m},x_{l}$ of which are included into
the different groups of the arguments. For the free field operators
$u_{\alpha (m)}(x_{m}),u_{\alpha (l)}(x_{l})$ the arguments
$x_{m},x_{l}$ of which are included into the same group the normal
product is supposed. For the chronological product we consider the
operators $:u_{\alpha (1)} (x_{1}) \cdots u_{\alpha (n_{i})}
(x_{n_{i}}):$, $i = 1,...,k$ as the whole objects.

Let the groups of the time arguments
$x_{1}^{0},...,x_{n_{1}}^{0};x_{n_{1} + 1}^{0},...,x_{n_{1} +
n_{2}}^{0};...;x_{n_{1} + \cdots n_{k - 1} + 1}^{0},...,$

\noindent $x_{n_{1} + \cdots n_{k}}^{0}$ are ordered due to the
subdivision $i_{1},...,i_{l}$, $j_{1},...,j_{k - l}$ of the numbers
$1,...,k$: any argument from the first group exceeds any argument
from the second group $x_{m}^{0} > x_{q}^{0}$ for any numbers $n_{1}
+ \cdots + n_{i_{s} - 1} < m \leq n_{1} + \cdots + n_{i_{s}}$, $s =
1,...,l$, and $n_{1} + \cdots + n_{j_{t} - 1} < q \leq n_{1} +
\cdots + n_{j_{t}}$, $t = 1,..,k - l$. By making use of the
definition of the mixed chronological and normal product of the free
field operators it is easy to prove the following relation
\begin{eqnarray}
\label{2.22} T\left( :\prod_{s\, =\, 1}^{n_{1}} u_{\alpha
(s)}(x_{s}):;\cdots ;:\prod_{s\, =\, n_{1} + \cdots + n_{k - 1} +
1}^{n_{1} + \cdots + n_{k}} u_{\alpha (s)}(x_{s}):\right) = \nonumber \\
(- 1)^{p}T\left( :\prod_{m\, =\, n_{1} + \cdots + n_{i_{1} - 1} +
1}^{n_{1} + \cdots + n_{i_{1}}} u_{\alpha (m)} (x_{m}):;\cdots
;:\prod_{m\, =\, n_{1} + \cdots + n_{i_{l} - 1} + 1}^{n_{1} + \cdots
+ n_{i_{l}}}
u_{\alpha (m)} (x_{m}):\right) \times \nonumber \\
T\left( :\prod_{q\, =\, n_{1} + \cdots + n_{j_{1} - 1} + 1}^{n_{1} +
\cdots + n_{j_{1}}} u_{\alpha (q)} (x_{q}):;\cdots ;:\prod_{q\, =\,
n_{1} + \cdots + n_{j_{k - l} - 1} + 1}^{n_{1} + \cdots + n_{j_{k -
l}}} u_{\alpha (q)}(x_{q}): \right)
\end{eqnarray}
where $p$ is the parity of the Fermi operators permutation. The
relation (\ref{2.22}) for the mixed chronological and normal product
of the free quantum fields is probably the differential form of the
causality condition.

\section{Scattering Matrix}
\setcounter{equation}{0}

Let us seek for a scattering matrix in the form
\begin{eqnarray}
\label{3.10} S(h) = 1 + \sum_{k\, =\, 1}^{\infty}
\sum_{n_{1},...,n_{k}}  \sum_{\alpha (1),...,\alpha (n_{1} + \cdots
+ n_{k})} \nonumber \\ \frac{1}{k!} K_{\alpha (1),...,\alpha
(n_{1})} \cdots K_{\alpha (n_{1} + \cdots + n_{k - 1} +
1),...,\alpha (n_{1} + \cdots + n_{k})} \times \nonumber
\\ \int d^{4}x_{1} \cdots
d^{4}x_{n_{1} + \cdots + n_{k}} \left(
h_{n_{1}}(x_{1},...,x_{n_{1}}) \cdots h_{n_{k}}(x_{n_{1} + \cdots +
n_{k - 1} + 1},...,x_{n_{1} + \cdots + n_{k}})\right) \times \nonumber \\
T\left( :\prod_{i\, =\, 1}^{n_{1}} u_{\alpha (i)}(x_{i}):;\cdots
;:\prod_{i\, =\, n_{1} + \cdots + n_{k - 1} + 1}^{n_{1} + \cdots +
n_{k}} u_{\alpha (i)}(x_{i}):\right)
\end{eqnarray}
where the mixed chronological and normal product of free quantum
field operators is defined above; in order to guarantee the scalar
character of the normal product Fermi operators must be included in
the even combinations only; $K_{\alpha (1),...,\alpha (n)}$ are the
constants; the switching functions $h_{n}(x_{1},...,x_{n}) \in
D({\bf R}^{4n})$. We consider that the natural numbers
$n_{1},...,n_{k}$ and the indexes $\alpha (1),...,\alpha (n_{1} +
\cdots + n_{k})$ in the equality (\ref{3.10}) run over the finite
sets of values.

If we insert into the equality (\ref{3.10}) the distributions
\begin{equation}
\label{3.11} h_{n}(x_{1},...,x_{n}) = g(x_{1})\delta (x_{2} -
x_{1})\cdots \delta (x_{n} - x_{1}),
\end{equation}
we get the expression similar to the scattering matrix expression in
the paper (\cite{8}, relations (5), (17)) and in the book (\cite{1},
relation (18.32)). Bogoliubov${}^{2}$ also believed that for the
physical scattering matrix the switching function $g(x)$ is equal to
$1$ in the relations (\ref{3.11}). Substituting the distributions
(\ref{3.11}) into the operator (\ref{3.10}) we have the diverging
integrals: the distributions should be integrated with the smooth
functions rapidly decreasing at the infinity.

In the book (\cite{1}, relation (18.27)) the chronological product
of the local operators is defined by means of the relation analogous
to the relation (\ref{2.7}). Due to the paper \cite{8}:

\noindent "Let us note as Stueckelberg did that the usual definition
of $T$ - product by means of introduction the chronological order
for the operators is effective only without the coincidence of the
arguments $x_{1},...,x_{n}$. In view of the corresponding
coefficient functions singularity their "redefinition" in the
domains of the arguments coincidence is not done explicitly and
presents a special problem...

If we do not call attention to this difficulty and use the Wick
theorem formally, then we get the expressions of the form:
\begin{equation}
\label{3.111} \prod_{a\, <\, b} D_{m_{ab}^{2}}^{c}(x_{a} - x_{b})
\end{equation}
consisting of the causal $D^{c}$ - functions products.

If we consider Fourier transform, then we get the integrals with the
well-known

\noindent "ultraviolet" divergences."

The local interaction Lagrangian in the scattering matrix
(\ref{3.10}), (\ref{3.11}) implies the singularities (\ref{3.111}).
The local interaction Lagrangian (\ref{2.61}) is the asymptotic
value of the polylocal normal product (\ref{2.60}): if the smooth
function $h(x_{1},..,x_{n})$ tends to the distribution
$g(x_{1})\delta (x_{2} - x_{1}) \cdots \delta (x_{n} - x_{1})$ where
the function $g(x_{1}) \in D({\bf R}^{4})$, then the polylocal
normal product (\ref{2.60}) tends to the local normal product
(\ref{2.61}). The chronological product for the local operators
(\ref{2.61}) is not correct. The interaction propagates not
instantly but at the speed not exceeding the speed of light. We have
to take into account the distance between the interacting particles.
Every interacting particle needs its own delay. We need to consider
the chronological product for the polylocal normal products
(\ref{2.60}).

Let us consider the scattering matrix (\ref{3.10}) with the
switching functions
\begin{equation}
\label{3.112} h_{n}(x_{1},...,x_{n}) = h_{n}^{(1)}(x_{1},...,x_{n})
+ h_{n}^{(2)}(x_{1},...,x_{n}).
\end{equation}
The support of the function $h_{n}^{(i)}(x_{1},...,x_{n})$ lies in
the domain $G_{i}^{\times n}$, $i = 1,2$, and all time points of the
domain $G_{2}$ lie in the future relative to all time points of the
domain $G_{1}$. The decomposition (\ref{3.112}) is analogous to the
decomposition (\ref{3.6}).

Let the subdivision $i_{1},...,i_{l}$, $j_{1},...,j_{k - l}$ of the
numbers $1,...,k$ be given. The relation (\ref{2.22}) implies
\begin{eqnarray}
\label{3.12} \int d^{4}x_{1} \cdots d^{4}x_{n_{1} + \cdots + n_{k}}
\left( \prod_{s\, =\, 1}^{l} h_{n_{i_{s}}}^{(2)}(x_{n_{1} + \cdots +
n_{i_{s} - 1} + 1},...,x_{n_{1} + \cdots + n_{i_{s}}})\right) \times
\nonumber
\\ \left( \prod_{t\, =\, 1}^{k\, -\, l} h_{n_{j_{t}}}^{(1)}(x_{n_{1} + \cdots +
n_{j_{t} - 1} + 1},...,x_{n_{1} + \cdots + n_{j_{t}}})\right) \times
\nonumber
\\ T\left( :\prod_{i\, =\, 1}^{n_{1}} u_{\alpha (i)}(x_{i}):;\cdots
;:\prod_{i\, =\, n_{1} + \cdots + n_{k - 1} + 1}^{n_{1} + \cdots +
n_{k}} u_{\alpha (i)}(x_{i}):\right) = \nonumber
\\ \int d^{4}x_{1} \cdots d^{4}x_{n_{1} + \cdots + n_{k}}
\left( \prod_{s\, =\, 1}^{l} h_{n_{i_{s}}}^{(2)}(x_{n_{1} + \cdots +
n_{i_{s} - 1} + 1},...,x_{n_{1} + \cdots + n_{i_{s}}})\right) \times
\nonumber
\\ \left( \prod_{t\, =\, 1}^{k\, -\, l} h_{n_{j_{t}}}^{(1)}(x_{n_{1} + \cdots +
n_{j_{t} - 1} + 1},...,x_{n_{1} + \cdots + n_{j_{t}}})\right) \times
\nonumber
\\ T\left( :\prod_{m\, =\, n_{1} + \cdots + n_{i_{1} - 1} + 1}^{n_{1} +
\cdots + n_{i_{1}}} u_{\alpha (m)} (x_{m}):;\cdots ;:\prod_{m\, =\,
n_{1} + \cdots + n_{i_{l} - 1} + 1}^{n_{1} + \cdots + n_{i_{l}}}
u_{\alpha (m)} (x_{m}):\right) \times \nonumber \\
T\left( :\prod_{q\, =\, n_{1} + \cdots + n_{j_{1} - 1} + 1}^{n_{1} +
\cdots + n_{j_{1}}} u_{\alpha (q)} (x_{q}):;\cdots ;:\prod_{q\, =\,
n_{1} + \cdots + n_{j_{k - l} - 1} + 1}^{n_{1} + \cdots + n_{j_{k -
l}}} u_{\alpha (q)}(x_{q}): \right).
\end{eqnarray}
The relations (\ref{3.12}) imply the equality
\begin{eqnarray}
\label{3.13} \sum_{n_{1},...,n_{k}} \sum_{\alpha (1),...,\alpha
(n_{1} + \cdots + n_{k})} \int
d^{4}x_{1} \cdots d^{4}x_{n_{1} + \cdots + n_{k}} \nonumber \\
\left( \prod_{s\, =\, 1}^{k} K_{\alpha (n_{1} + \cdots + n_{s - 1} +
1),...,\alpha (n_{1} + \cdots + n_{s})} h_{n_{s}}(x_{n_{1} + \cdots
n_{s - 1} + 1},...,x_{n_{1} + \cdots n_{s}})\right) \times \nonumber
\\ T\left( :\prod_{i\, =\, 1}^{n_{1}} u_{\alpha (i)}(x_{i}):;\cdots
;:\prod_{i\, =\, n_{1} + \cdots + n_{k - 1} + 1}^{n_{1} + \cdots +
n_{k}} u_{\alpha (i)}(x_{i}):\right) = \nonumber \\ \sum_{l\, =\,
0}^{k} \frac{k!}{l!(k - l)!} \Biggl\{ \sum_{n_{1},...,n_{l}}
\sum_{\alpha (1),...,\alpha (n_{1} + \cdots + n_{l})} \int
d^{4}x_{1} \cdots d^{4}x_{n_{1} + \cdots + n_{l}} \nonumber \\
\left( \prod_{s\, =\, 1}^{l} K_{\alpha (n_{1} + \cdots + n_{s - 1} +
1),...,\alpha (n_{1} + \cdots + n_{s})} h_{n_{s}}^{(2)}(x_{n_{1} +
\cdots n_{s - 1} + 1},...,x_{n_{1} + \cdots n_{s}})\right) \times
\nonumber
\\ T\left( :\prod_{i\, =\, 1}^{n_{1}} u_{\alpha (i)}(x_{i}):;\cdots
;:\prod_{i\, =\, n_{1} + \cdots + n_{l - 1} + 1}^{n_{1} + \cdots +
n_{l}} u_{\alpha (i)}(x_{i}):\right) \Biggr\} \times \nonumber
\\ \Biggl\{ \sum_{n_{1},...,n_{k - l}}
\sum_{\alpha (1),...,\alpha (n_{1} + \cdots + n_{k - l})} \int
d^{4}y_{1} \cdots d^{4}y_{n_{1} + \cdots + n_{k - l}} \nonumber \\
\left( \prod_{s\, =\, 1}^{k - l} K_{\alpha (n_{1} + \cdots + n_{s -
1} + 1),...,\alpha (n_{1} + \cdots + n_{s})}
h_{n_{s}}^{(1)}(y_{n_{1} + \cdots n_{s - 1} + 1},...,y_{n_{1} +
\cdots n_{s}})\right) \times \nonumber
\\ T\left( :\prod_{i\, =\, 1}^{n_{1}} u_{\alpha (i)}(y_{i}):;\cdots
;:\prod_{i\, =\, n_{1} + \cdots + n_{k - l - 1} + 1}^{n_{1} + \cdots
+ n_{k - l}} u_{\alpha (i)}(y_{i}):\right) \Biggr\}.
\end{eqnarray}
Inserting the equalities (\ref{3.13}) into the right-hand side of
the equality (\ref{3.10}) we get the equality
\begin{equation}
\label{3.14} S(h^{(1)} + h^{(2)}) = S(h^{(2)})S(h^{(1)})
\end{equation}
similar to the equality (\ref{3.9}).

It is possible to choose the constants in the equality (\ref{3.10})
such that the relation analogous to the relation (\ref{3.4}) is
valid
\begin{equation}
\label{3.15} S(Lh) = U(L)S(h)U^{\ast}(L).
\end{equation}
Here $Lh_{n}(x_{1},...,x_{n}) = h_{n}(L^{- 1}x_{1},...,L^{-
1}x_{n})$ and $U(L)$ is a unitary operator by means of which the
quantum wave functions transform under the transformations $L$ from
the Lorentz group and the group of translations.

In order to conserve the amplitude state norm under the
transformation from the initial state to the final state the
operator $S(h)S^{\ast}(h)$ has to be the identity operator.

\end{document}